\documentclass{article}[12pt]

\usepackage{a4,amsmath,amsthm,amssymb,amsfonts}
\usepackage{graphicx}
\usepackage{dcolumn}
\usepackage{bm}

\newcommand{\R}{{\mathbb R}}

\newcommand \h {\mathcal{H}}
\newcommand \ra {\rightarrow}
\newcommand{\ahalf}{\mbox{$\frac12$}}

\newcommand \I   {\mathrm{i}}
\newcommand{\be}{\begin{equation}}
\newcommand{\ee}{\end{equation}}
\newcommand   \eps \varepsilon

\theoremstyle{plain}
\newtheorem  {theorem}{Theorem}[section]

\newtheorem  {corollary}[theorem]   {Corollary}
\newtheorem* {corollary*}           {Corollary}

\begin{document}


\title{On the Ehrenfest theorem of quantum mechanics}

\author{{Gero Friesecke and Mario Koppen\footnote{
Zentrum Mathematik, Technische Universit\"at M\"unchen, Boltzmannstrasse 3, 85748 Garching,
Germany, gf@ma.tum.de, mario@ma.tum.de}}}

\date{March 28, 2009}


\maketitle

\begin{abstract}
We give a mathematically rigorous derivation of Ehrenfest's equations for the
evolution of position and momentum expectation values, under general and natural assumptions
which include atomic and molecular Hamiltonians with Coulomb interactions. 
\end{abstract}

\section{\label{sec:level1}INTRODUCTION}
A basic result in the physics literature asserts that the mean position and momentum of a quantum
system in $\R^d$ with Hamiltonian 
\be \label{ham}
    H=-\sum_{j=1}^d \frac{1}{2m_j}\frac{\partial^2}{\partial x_j^2}+ V(x)
\ee 
evolve ``classically'', 
\begin{eqnarray} \label{eh1}
   & & \frac{d}{dt} \langle X_j \rangle_{\psi(t)} = \frac{1}{m_j} \langle P_j \rangle_{\psi(t)}, \\
   & & \frac{d}{dt} \langle P_j \rangle_{\psi(t)} = \langle -\frac{\partial V}{\partial x_j} \rangle_{\psi(t)} \label{eh2}
\end{eqnarray}
(Ehrenfest's equations, \cite{Ehrenfest}). Here $\psi(t)\in L^2(\R^d)$ denotes the state of the quantum system at 
time $t$, assumed to evolve via the time-dependent Schr\"odinger equation 
\be \label{TSE}
    i\frac{d}{dt}\psi(t)=H\psi(t),
\ee 
$\langle A \rangle_{\psi(t)}$ stands for the mean (or `expected value') $\langle\psi(t),\, A\psi(t)\rangle$ of a linear
operator (or `observable') $A$ on $L^2(\R^d)$, $\langle f, \, g\rangle=\int_{\R^d} \overline{f} \, g$ 
is the usual $L^2$ inner product, $X$ and $P$ are the position and momentum operators (see (\ref{ham2}), 
(\ref{ham3}) below), and the potential $V \, : \, \R^d\to\R$ and the force $-\nabla V\, : \, \R^d\to\R^d$ 
act by multiplication. 
The evolution of mean values as in (\ref{eh1}), (\ref{eh2}), 
besides being of interest in its own right, 
plays an important role in the study of quantum-classical 
coupling in molecular dynamics (see e.g. \cite{Bornemannetal, Griebeletal}). Also, a priori estimates related to
\eqref{eh1}, \eqref{eh2} are utilized in the proofs of important results in scattering theory \cite{Sigal, Graf,
Derezinski, DG}.

The heuristic justification, which can be found in any quantum mechanics textbook, goes as follows. 
When $H$ and $A$ are hermitean, formally differentiating the mean of $A$
and substituting into the Schr\"odinger equation yields 
\be \label{eh3}
      \frac{d}{dt}\langle A\rangle_{\psi(t)}=i\langle [H,A]\rangle_{\psi(t)},
\end{equation}
where $[H,A]$ denotes the commutator $HA-AH$. When $H$ is of the form (\ref{ham})
and $A$ is a component of the position or momentum operator, formal evaluation of the commutator
gives (\ref{eh1}), (\ref{eh2}). For interesting discussions in the physics literature on the shortcomings
of the formal argument see \cite{ADS1, ADS2, ADS3, Hill}. Also, it is straightforward
to make Ehrenfest type equations rigorous
under sufficiently stringent assumptions, such as: $H$ is self-adjoint and $A$ is bounded and leaves the domain of $H$ invariant; or: 
$H$ is a Schr\"odinger operator with smooth potential, $A$ is a component of the position or momentum operator, and $\psi(t)$ belongs
to Schwartz space for all $t$, the Schwartz norms being uniformly bounded when $t$ belongs to a bounded interval.  
\\[2mm]
To our knowledge, a rigorous version of \eqref{eh1}, \eqref{eh2} under satisfactory assumptions, which 
include in particular the basic atomic and molecular Hamiltonians with Coulomb interactions, is so far missing from the literature.
It is however readily obtained by combining known results (in particular, \cite{Hunziker, RS}) with
standard functional analytic arguments, as we point out in this note. 

Making rigorous sense of the terms appearing in the abstract Ehrenfest equation (\ref{eh3}) 
when $H$ and $A$ are unbounded self-adjoint operators requires, in particular: 
\begin{itemize}
\item [(1)] Well-definedness of the expected value
$\langle \psi(t), \, A\psi(t)\rangle$ for all $t$. When the latter is interpreted as an inner product, this means
invariance of $D(A)$ under the propagator $e^{-itH}$ of eq. (\ref{TSE}). 
\item[(2)] Differentiability of $\langle \psi(t), \, A\psi(t)\rangle$ with respect to $t$.
\item[(3)] Well-definedness of the expected value of the commutator $HA-AH$ for all $t$, in some suitable sense.
\end{itemize}
As turns out, (1) is essentially sufficient for (2), (3), and the validity of eq. (\ref{eh3}):
\begin{theorem} (Abstract Ehrenfest theorem) \label{AbstractResult} Let $H$ and $A$ be two densely defined linear operators on a
Hilbert space $\h$ such that: 
\\[1mm]
{\rm (H1)} $H:D(H)\ra\h$ is selfadjoint, $A:D(A)\ra \h$ is hermitean \\
{\rm (H2)} ${\rm e}^{-\I tH}$ leaves $D(A)\cap D(H)$ invariant for all $t\in\R$ \\
{\rm (H3)} For any $\psi_0\in D(A)\cap D(H)$, $\sup_{t\in I}||Ae^{-itH}\psi_0||<\infty$ for $I\subset\R$ bounded.
\\[1mm]
Then for $\psi_0\in D(A)\cap D(H)$, the expected value $\langle A \rangle_{\psi(t)}$, $\psi(t):=e^{-itH}\psi_0$, is 
continuously differentiable with respect to $t$ and satisfies eq. (\ref{eh3}), the right hand side being defined as the
following quadratic form:
\be \label{formdef}
   \langle [H,A]\rangle_{\psi}:= \langle H\psi,A\psi\rangle-\langle A\psi,H\psi\rangle \;\; \mbox{($\psi\in D(A)\cap D(H)$)}.
\ee
\end{theorem}
{\bf Remarks} 1) Assumption (H2) is needed to make the expected value
$\langle A \rangle_{\psi(t)}$ well defined (see (1)), and (H3) will be useful in establishing
its differentiability (see (2) and the proof below). Note that (H2) and (H3) hold automatically when
$A$ is relatively bounded with respect to $H$. 
\\[2mm]
2) Definition (\ref{formdef}) allows us to make sense
of the right hand side of (\ref{eh3}) without the need to consider the (often complicated) domains of the
composite operators $AH$ and $HA$, let alone proving their invariance under $e^{-itH}$.   
Note that these domains need not be dense when $A$ and $H$ are general self-adjoint operators, and 
do not even contain $C_0^\infty(\R^d)$ in basic examples from physics (see below). Of course,
when $\psi$ belongs $D(AH)\cap D(HA)$ (where $D(AH):=\{\psi\in D(H)\, | \, H\psi\in D(A)\}$), (\ref{formdef}) reduces to the classical definition 
$\langle \psi, (HA-AH)\psi\rangle$.

{\bf Example} Let $H$ be the
hydrogen atom Hamiltonian and $A$ the momentum operator, 
$H=-\frac{1}{2}\Delta + V(x)$, $V(x)=-1/|x|$, $x\in\R^3$. On $C_0^\infty(\R^3)$ 
we have $i[H,A]=-\nabla V(x)=-x/|x|^3$. 
But $-x/|x|^3$ does not map $C_0^\infty(\R^3)$ into 
$L^2(\R^3)$, because the singularity at zero behaves like $|x|^{-2}$ and hence $||\nabla V \, \phi||_{L^2}^2 = 
\int_{\R^d}|\nabla V(x)\phi(x)|^2dx = \infty$ for all $\phi\in C_0^\infty(\R^3)$ with $\phi(0)\neq 0$. 
\\[2mm]
3) It is a deeper fact is that if $A$ is in addition assumed to be self-adjoint then {\rm (H1)} and {\rm (H2)} imply {\rm (H3)}.
A proof of this fact, as well as a counterexample showing that hermiteanity does not suffice for this implication, will
appear elsewhere \cite{FS}.
\\[2mm]
4) The main point in the proof is a weak convergence argument which exploits the bound in (H3); see Section \ref{SecAbstract}. 
\\[2mm]
Next, we describe our ensuing results on equations \eqref{eh1}, \eqref{eh2}. Let $H$ be a Hamiltonian of the form
\begin{equation} \label{ham1}
   H=T+V(x),\;T=-\sum_{j=1}^d \frac{1}{2m_j}\frac{\partial^2}{\partial x_j^2},\;
    m_1,..,m_d>0, \; \h=L^2(\R^d), \; D(H)=H^2(\R^d), 
\end{equation}
(for conditions on $V$ such that $H$ is well defined and self-adjoint on $H^2(\R^d)$ see below)
 and let $A$ 
a component of the position or momentum operator, 
\begin{equation}
  (X_j\psi)(x)=x_j\psi(x),\;D(X_j)=\{\psi\in L^2(\R^d) \, | \, \int_{\R^d} |x_j|^2\; |\psi(x)|^2 dx <\infty\}, \label{ham2}
\end{equation}
\begin{equation} \label{ham3}
   P_j=\frac{1}{i} \frac{\partial}{\partial x_j}, \;\; D(P_j)=\{\psi\in L^2(\R^d)\, | \, p_j\widehat{\psi}(p)\in L^2(\R^d)\},
\end{equation}
(where $\widehat{\psi}$ denotes the Fourier transform of $\psi$, defined on smooth rapidly decaying functions
by $\widehat{\psi}(p)=\int_{\R^d}e^{-ip\cdot x}\psi(x)\, dx$). Clearly, $X_j$ and $P_j$ are hermitean.
In order to guarantee well-definedness and self-adjointness of $H$ on $D(H)$, by the Kato-Rellich theorem
it suffices to assume that $V \, : \, \R^d\to\R$ is a real-valued locally integrable function such that
\begin{equation}
   \psi\mapsto V\psi \mbox{ is relatively bounded w.r.t.}\,
   T\mbox{ with relative bound }\alpha<1.\label{katorellich}
\end{equation}
By this one means that for all $\psi\in H^2(\R^d)$ the function $V\psi$ belongs to $L^2(\R^d)$ and 
$||V\psi||\le \alpha ||T\psi||+C_\alpha||\psi||$, for some constants $\alpha<1$ and $C_\alpha$. 
Prototypical are the electronic Hamiltonian of a general molecule,
$d=3N$, $x=(y_1,..,y_N)$, $y_j\in\R^3$, 
\begin{equation} \label{Helec}
  H_{e\ell} = -\ahalf \Delta_y + \sum_{i=1}^N v(y_i) + \sum_{1\le i<j\le N}\frac{1}{|y_i-y_j|},
  \quad 
  v(y_i)=-\sum_{\alpha=M}\frac{Z_\alpha}{|y_i-R_\alpha|},\nonumber
\end{equation}
and the full electron-nuclei Hamiltonian of such a molecule, $d=3(N+M)$, $x=(y_1,..,y_N,R_1,..,R_M)$, $y_i\in\R^3$, $R_\alpha\in\R^3$,
\be \label{Hmolec}
  H = H_{e\ell} -\sum_{\alpha=1}^M\frac{1}{2M_\alpha}\Delta_{R_{\alpha}} + \sum_{1\le \alpha<\beta\le M}\frac{Z_\alpha
  Z_\beta}{|R_\alpha-R_\beta|}.
\ee
Here the $M_\alpha>0$ and $Z_\alpha>0$ are the masses and charges of the nuclei, and atomic units have been used so that
$\hbar=1$ and the electrons have mass $1$ and charge $-1$.
\begin{corollary} \label{weakehrenfest} (Time evolution of expected position and momentum) \\
Let $H$ be given by (\ref{ham1}), with $V \, : \, \R^d\to\R$ being a real-valued locally integrable function
satisfying the Kato-Rellich condition (\ref{katorellich}). Then:
\begin{itemize}
\item[(i)] $\langle X_j\rangle_{\psi(t)}$ is continuously differentiable with respect to $t$ for any $\psi_0\in
D(X_j)\cap D(H)$, and satisfies the equation 
$$
     \frac{d}{dt}\langle X_j\rangle_{\psi(t)} = i \Bigl(\langle H\psi(t), X_j\psi(t)\rangle - \langle X_j\psi(t), H\psi(t)\rangle\Bigr).
$$
\item[(ii)] $\langle P_j\rangle_{\psi(t)}$ is continuously differentiable with respect to $t$ for any $\psi_0\in D(H)$, and satisfies the equation 
$$
     \frac{d}{dt}\langle P_j\rangle_{\psi(t)} = i \Bigl(\langle H\psi(t), P_j\psi(t)\rangle - \langle P_j\psi(t), H\psi(t)\rangle\Bigr).
$$
\end{itemize}
\end{corollary}
{\bf Proof} It suffices to check that the hypotheses of Theorem \ref{AbstractResult} are satisfied. Clearly, (H1) holds for both $X_j$ and $P_j$.
In the case of $X_j$, (H2) and (H3) are satisfied by the results of \cite{Hunziker, RS}. As regards $P_j$, 
(H2) and (H3) are satisfied since $D(H)\cap D(P_j)=D(H)$, $P_j$ is relatively bounded
with respect to $H$ (i.e. $||P_j\psi||\le C(||H\psi||+||\psi||)$ for all $\psi\in D(H)$),
as is easily deduced from the Kato-Rellich condition, and $||\psi(t)||$ and $||H\psi(t)||$ are time-invariant.
\\[2mm]
Note, however, that so far we have not fully derived the classical Ehrenfest equations for position and momentum, since it remains
to be verified that the right hand sides in (i) and (ii) agree with the classical right hand sides in (\ref{eh1}), (\ref{eh2}). 
In particular, in order to recover the classical expression $\langle -\frac{\partial V}{\partial x_j}\rangle_{\psi(t)}$, i.e. 
the quadratic form of the multiplication operator corresponding to the $j^{th}$ component of the force, additional assumptions on
the potential are needed. 
\begin{theorem} \label{PosMomThm}
Let $H$ be given by (\ref{ham1}), where $V \, : \, \R^d\to\R$ is real-valued, belongs to the Sobolev space $W^{1,1}_{\ell oc}(\R^d)$
of locally integrable functions with locally integrable weak derivatives, and $V$ and $\sqrt{|\nabla V|}$ satisfy 
the Kato-Rellich condition (\ref{katorellich}). Then:
\begin{itemize}
\item[(i)] $\langle X_j\rangle_{\psi(t)}$ is continuously differentiable with respect to $t$ for any $\psi_0\in
D(X_j)\cap D(H)$, and satisfies (\ref{eh1}).
\item[(ii)] $\langle P_j\rangle_{\psi(t)}$ is continuously differentiable with respect to $t$ for any $\psi_0\in D(H)$, and satisfies 
(\ref{eh2}).
\end{itemize}
\end{theorem}
An important technical point regarding the assumptions of Theorem \ref{PosMomThm} is that the Kato-Rellich condition is not required for $\nabla V$,
but only its square root. This is related to the fact that $\nabla V$ is only needed as a quadratic form, see
Sec. \ref{Sec:Classical}. Hence the Hamiltonians (\ref{Helec}), (\ref{Hmolec}) satisfy the assumptions of the theorem. This is because
the square root of the gradient of a typical term in $V$ looks the same as the term itself, 
$|\nabla_{y_i}|y_i-R_\alpha|^{-1}|^{1/2} = |y_i-R_\alpha|^{-1}$.  
Note that for these Hamiltonians, $\nabla V$ itself fails the Kato-Rellich condition (it does not even map $C_0^\infty$ to $L^2$, see the
example in Remark 2) above.

\section{Proof of abstract Ehrenfest theorem} \label{SecAbstract}
\begin{proof}
We begin by recalling the properties of the strongly continuous one-parameter unitary group $e^{-itH}$ generated
by a self-adjoint operator $H$ (see e.g. \cite{RSI}). For all $t$, $e^{-itH}$ leaves $D(H)$ invariant and commutes on $D(H)$ with $H$;
moreover for any $\psi_0\in D(H)$, $t\mapsto\psi(t)=e^{-itH}\psi_0$ is a continuously
differentiable map from $\R$ to $\h$ and satisfies
$i\frac{d}{dt}\psi(t)=H\psi(t)$ for all $t$.

In order to show the existence of 
\begin{align*}
   \frac{d}{dt}\langle \psi(t),A\psi(t)\rangle=&
   \underset{h\ra 0}{\lim}
   \frac{
    \langle \psi(t\!+\! h),A\psi(t\!+\! h)\rangle-\langle \psi(t), A\psi(t)\rangle
   }{h},
\end{align*}
we use the decomposition 
\[
    \frac{\langle \psi(t\!+\! h),\, A\psi(t\!+\! h)\rangle
    - \langle \psi(t), \, A\psi(t)\rangle}{h}
    =\Bigl\langle A\psi(t\!+\! h), \, \frac{\psi(t\!+\! h)-\psi(t)}{h} \Bigr\rangle
    +
    \Bigl\langle \frac{\psi(t\!+\! h)-\psi(t)}{h}, \, A\psi(t)\Bigr\rangle
\]
and show the existence of the limits of the two terms on the RHS separately. 
Since $\frac{1}{h}(\psi(t\!+\! h)-\psi(t)){\ra} -\I H \psi(t)$ strongly in $\h$ for $h\to 0$, the second term converges
to $i\langle H\psi(t), A\psi(t)\rangle$. Moreover, the first term goes to $-i\langle A\psi(t),H\psi(t)\rangle$ provided
we can show that $A\psi(t\!+\! h){\rightharpoonup} A\psi(t)$ weakly in $\h$ for $h\to 0$.
To this end, 
fix $t\in \R$ and choose a sequence $\{h_j\}\subset \R$ satisfying $h_j\ra 0$. Then
by (H3), the set $\{A\psi(t\!+\! h_j)\}$ is bounded in $\h$.
By weak compactness of the unit ball, there exists a subsequence (again labelled by $h_j$) such that
 $A\psi(t\!+\! h_j)\rightharpoonup f$ for some $f\in \h$. We claim that $f=A\psi(t)$. To see this, 
choose any $\phi\in D(A)$ and calculate using the weak convergence of $A\psi(t+h_j)$, the hermiteanity of $A$, the 
continuity of $\psi(t)$ in $t$ and the fact that
$\psi(t)\in D(A)$
\[
  \langle f, \, \phi\rangle =\lim_{h_j\to 0}\langle A\psi(t\!+\! h_j), \,\phi\rangle
  = \lim_{h_j\to 0}
  \langle\psi(t\!+\! h_j), A\phi\rangle
  = \langle \psi(t), A\phi \rangle
  = \langle A\psi(t), \phi\rangle.
\]
 Since $D(A)$ is dense in $\h$, the claim follows, and since the argument is valid for all 
 sequences $h_j\ra 0$, it follows that $A\psi(t\!+\! h)\rightharpoonup A\psi(t)$. This completes the proof of differentiability
of $\langle A \rangle_{\psi(t)}$ and of eq. (\ref{eh3}). It remains to show that the derivative $2 \, {\rm Im}\, \langle H\psi(t), \, A\psi(t)\rangle$ is continuous in $t$. Indeed, for $h\to 0$ we
have $H\psi(t\!+\! h)\to H\psi(t)$ strongly in $\h$ and, as just shown, $A\psi(t\!+\! h)\rightharpoonup A\psi(t)$ weakly in 
$\h$, completing the proof. 
\end{proof}
\section{Derivation of the classical Ehrenfest equations} \label{Sec:Classical}

{\bf Proof of Theorem \ref{PosMomThm} (i)}
Thanks to Corollary \ref{weakehrenfest}, we only need to evaluate the abstract expected value
(\ref{formdef}) of the commutator and show that 
\begin{equation} \label{formdefXj}
   i\Bigl( \langle H\psi, X_j\psi\rangle - \langle X_j\psi, H\psi\rangle\Bigr)
   = \frac{1}{m_j}\, \langle\psi, \, P_j\psi\rangle
   \mbox{ for all }\psi\in H^2(\R^d)\cap D(X_j).
\end{equation}
For functions $\psi\in C_0^\infty(\R^d)$ this follows from an elementary calculation. The general case will follow from 
an approximation argument, but due to the presence of the unbounded operator $X_j$, a little care is needed. 

First, consider functions $\psi\in H^2(\R^d)$ with compact support. Clearly it suffices to approximate $\psi$ by 
a sequence $\psi_\eps$ of $C_0^\infty$ functions in such a way that the four terms appearing inside the inner products,
$H\psi_\eps$, $X_j\psi_\eps$, $P_j\psi_\eps$ and $\psi_\eps$, converge in $L^2$ to the corresponding terms for $\psi$. 
Choose a ball $B_R(0)$ containing the support of $\psi$. Consider the following standard approximation obtained by mollification:
$\psi_\eps(x)=(\chi_\eps*\psi)(x)=\int_{\R^d}\chi_\eps(x-y)\psi(y)dy$, where $\chi_\eps(x)=\eps^{-d}\chi(x/\eps)$, $\eps\in(0,1)$,
$\chi\in C^\infty_0(\R^d)$, $\chi=0$ outside $B_1(0)$, $\int_{\R^d}\chi=1$. Then (see e.g. \cite{Evans}) $\psi_\eps\in C_0^\infty$,
$\psi_\eps\to \psi$ in $H^2$. Since $H$, $P_j$ and $I$ are continuous operators from $H^2$ to $L^2$ (in case of $H$ this follows
from (\ref{katorellich})), $H\psi_\eps\to H\psi$, $P_j\psi_\eps\to P_j\psi$ and $\psi_\eps\to\psi$ in $L^2$. Finally, supp$\,\psi_\eps$, 
supp$\,\psi\subset B_{R+1}(0)$, and $X_j$ is bounded on the subspace of $L^2$ functions with support in $B_{R+1}(0)$, so $X_j\psi_\eps\to
X_j\psi$ in $L^2$. This establishes (\ref{formdef}) for compactly supported $H^2$ functions.

Finally let $\psi$ be a general function in $D(H)\cap D(X_j)=H^2\cap D(X_j)$. 
Let $\chi\in C_0^\infty$ with $\chi(0)=1$. Then $\psi_R(x):=\chi(x/R)\psi(x)$ is a compactly supported $H^2$ function, so
(\ref{formdefXj}) holds for $\psi_R$ by the previous step, 
and it is straightforward to check that
$\psi_R\to\psi$ in $H^2$. Moreover $X_j\psi_R$ converges to $X_j\psi$ in $L^2$
since $X_j\psi_R=(X_j\psi)_R$ and $X_j\psi\in L^2$. Consequently $H\psi_R$, $X_j\psi_R$, $P_j\psi_R$ and $\psi_R$ converge in $L^2$ to the corresponding 
terms for $\psi$, establishing (\ref{formdefXj}) in the general case.
\\[2mm]
{\bf Proof of Theorem \ref{PosMomThm} (ii)}
As in the proof of (i), it only remains for us to evaluate the abstract expected value
(\ref{formdef}) of the commutator and to show that 
\be \label{Pjcomm}
   \underbrace{i\Bigl( \langle H\psi, P_j\psi\rangle - \langle P_j\psi, H\psi\rangle\Bigr)}_{=:Q(\psi)}
  = \underbrace{\langle \psi, -\mbox{$\frac{\partial V}{\partial x_j}$}\psi\rangle}_{=:Q_*(\psi)}, 
\ee
for all $\psi\in H^2(\R^d)$. We start by considering $\psi\in C_0^\infty(\R^d)$. In this case, an elementary
calculation shows that
\be 
  Q(\psi) = \int_{\R^d}V(x)\mbox{$\frac{\partial}{\partial x_j}$}|\psi(x)|^2 dx.
\ee
Due to the assumption $V\in W^{1,1}_{\ell oc}(\R^d)$ and the fact that $|\psi(x)|^2\in C_0^\infty(\R^d)$, 
we can integrate by parts to obtain (\ref{Pjcomm}) for all $\psi\in C_0^\infty(\R^d)$. 

To establish (\ref{Pjcomm}) for all $\psi\in H^2$, by the density of $C_0^\infty$ in $H^2$ it suffices to show
that the quadratic forms $Q$ and $Q_*$ are continuous on $H^2$. As regards $Q$, this follows from the fact 
that $H$ and $P_j$ are continuous operators from $H^2$ to $L^2$. As regards $Q_*$, this follows by writing 
$$
   Q_*(\psi) = \langle f\psi, \, g\psi\rangle, \;\; f:=\sqrt{|\mbox{$\frac{\partial V}{\partial x_j}$}|}, \;\; 
                                                  g:= sgn(\mbox{$\frac{\partial V}{\partial x_j}$}) \, 
                                                  \sqrt{|\mbox{$\frac{\partial V}{\partial x_j}$}|}
$$
and noting that $\psi\mapsto f\psi$, $\psi\mapsto g\psi$ are continuous operators from $H^2$ to $L^2$ by 
the relative boundedness of $f$ with respect to $T$ and the fact that $T\, : \, H^2\to L^2$ is continuous. 
The proof of Theorem \ref{PosMomThm} is complete.

\bibliographystyle{alpha}
\bibliography{ehrenaps}

\end{document}